# On the Performance of Machine Learning Methods for Breakthrough Curve Prediction*

Daria Fokina[1,2], Oleg Iliev[1,2,3], Pavel Toktaliev[1,2], Ivan Oseledets[4], Felix Schindler[5]

[1]Fraunhofer ITWM, Kaiserslautern, Germany

[2]Technische Universität Kaiserslautern, Kaiserslautern, Germany

[3]Institute of Mathematics and Informatics, Bulgarian Academy of Sciences, Sofia, Bulgaria

[4]Skolkovo Institute of Science and Technology, Moscow, Russia

[5]Mathematics Münster, Westfälische Wilhelms-Universität Münster, Germany

**Summary:**

Reactive flows are important part of numerous technical and environmental processes. Often monitoring the flow and species concentrations within the domain is not possible or is expensive, in contrast, outlet concentration is straightforward to measure. In connection with reactive flows in porous media, the term *"breakthrough curve"* is used to denote the time dependency of the outlet concentration with prescribed conditions at the inlet. In this work we apply several machine learning methods to predict breakthrough curves from the given set of parameters. In our case the parameters are the Damköhler and Peclet numbers. We perform a thorough analysis for the one-dimensional case and also provide the results for the three-dimensional case.

**Keywords:**

Machine learning, Breakthrough curve prediction, Catalytic filters, Filtration, Separation, Purification,

*Funded by BMBF under contracts 05M20AMD and 05M20PMA. Funded by the Deutsche Forschungsgemeinschaft (DFG, German Research Foundation) under Germany's Excellence Strategy EXC 2044 - 390685587, Mathematics Münster: Dynamics – Geometry – Structure. Ivan Oseledets appreciates funding via Alexander von Humboldt Research Award.

# 1   Introduction

Filtration, purification and separation play a critical role in many industrial, environmental and biomedical processes. Finding advanced solutions in these areas is important for the development of highly efficient and reliable products and tools, as well as for ensuring a high quality of life for the general public. It is difficult to find an industry or area of life where filters do not play an important role. In one only car, for example, there are filters for the exhaust gases, transmission, cabin air, fuel, engine air, coolant, and brake systems. Furthermore, the quality of our drinking water, the treatment of wastewater, the air we breathe at home or in the office—everything is critically dependent on filtration solutions. The filtration and separation business is expanding rapidly, with scores of large companies and thousands of SMEs competing to develop better filters. The industrial demand for innovative filtration and purification solutions is growing steadily, thus promoting the use of Computer Aided Engineering in designing filter media and filter elements, in finding optimal filtration, separation and purification solutions. The so-called breakthrough curves often provide a way to measure the efficiency of filtration and separation processes. In the particular case of catalytic filters for exhaust gases, the term conversion rate is used. The optimal design of efficient filtration and separation equipment, the optimal control of the filtration and separation processes, the parameter identification of unknown reaction rates, usually require the knowledge of a large number of breakthrough curves. Computer Aided Engineering can assist optimization tasks by reducing the number of required measurements and replacing the rest of the measurements with computer simulations. Till recently direct computer simulations were used for this purpose, i.e., solving systems of partial differential equations. Recently it was found that in many cases machine learning can be more efficient compared to experiments and solving systems of PDEs.

Methods of machine learning, ML, are being actively developed and widely used nowadays. Most state of the art models, e.g. transformers, are specifically designed for computer vision and natural language processing tasks. These models require a huge amount of data. In areas, where it is currently impossible to collect that amount of data, simpler models are used. For example, Gaussian processes have successfully found an application in several areas, e.g. in earthquake engineering [6] and topology optimization [7]. Physics-informed neural networks [5] are an example of a small neural network used as a partial differential equation solver.

In this work we focus on the usage of machine learning methods to assist optimization of catalytic filters. The physical process of exhaust gases purification can be described by a convection-diffusion-reaction partial differential equation(s). The efficiency of the filter (called conversion rate in the case of catalytic filters) can be described by the time-dependent concentration of impurities on the outlet, which is called *a breakthrough curve*. To assist in the optimization of catalytic filters (e.g. selection of active material - washcoat with proper reactivity or selection of most appropriate flow regime) or in parameter identification problems (aiming at identifying unknown washcoat reactivities), one needs to evaluate a large number of breakthrough curves, corresponding to different process parameters. As mentioned above, machine learning could help to do this efficiently. To estimate the breakthrough curve via machine learning, one needs an initial set of breakthrough curves obtained either from measurements or by solving the respective system of PDEs. After a training stage, machine learning algorithms provide a fast way to predict breakthrough curves directly as a function of process parameters (in terms of the adopted mathematical model, they are also input parameters of the governing PDEs). In this case ML methods as surrogate models can be much more efficient than classical direct approaches. The only requirement is a sufficient number of training samples for the ML model. Here we should note that we predict the breakthrough curve not as a continuous function in time but as a vector of function values at particular moments of time.

We consider Gaussian processes and neural networks machine learning methods. Further, we also consider the application of tensor approximation for the given task.

# 2   Problem formulation

In this work we consider a reactive flow of gas mixture through filter media with a predefined velocity field. We assume that contaminant concentration, $c \in \mathrm{R}$, is small enough to neglect back coupling with the bulk flow and the species transport problem can be decoupled from the flow problem. Species transport can be described by a parametric convection-diffusion-reaction PDE(s). For the case of one linear reaction and one gaseous contaminant in a non-dimensionalized form it is written as:

$$\frac{\partial c}{\partial t} - \Delta c + Pe \cdot (uc) + Da \cdot c = 0, c = c(x,t), \quad x \in \Omega, 0 < t \leq T \tag{1}$$

$$\text{BC: } c(x,t) = 1, \quad x \in \Omega_{\text{inlet}} \tag{2}$$

$$\frac{\partial c}{\partial x}(x,t) = 0, \quad x \in \Omega_{\text{outlet}} \tag{3}$$

$$\text{IC: } c(x, 0) = 0 \tag{4}$$

where $Da > 0$ is the Damköhler number, $Pe > 0$ is the Peclet number and u is the predefined velocity field. Here we use characteristic diffusion time as a base for dimensionless form and time, T, large enough to reach the equilibrium. The domain Ω and its subsets $Ω_{inlet}$, $Ω_{outlet}$ represents filter media, inlet and outlet section respectively. We aim to predict for a given pair of coefficients $μ = (Da, Pe)$ the breakthrough curve:

$$s_μ(t) = \int_{Ω_{outlet}} c(x,t)dx. \tag{5}$$

Instead of taking $s(t)$ as a continuous function in time, we consider only values of $s_μ(t)$ in fixed moments of time uniformly distributed over the interval $[0,T]$ and work with vectors $s(μ) := s_μ(t) \in \mathbb{R}^{N_T}$, $t = [0, Δt, 2Δt, \ldots, T]$. Depending on ranges $Da$ and $Pe$ values, we consider the following regimes:

- Diffusion dominated: $Da \in [10^{-2}, 1], Pe \in [10^{-2}, 1]$
- Convection dominated, small $Da$: $Da \in [10^{-2}, 1], Pe \in [5, 10^2]$

For each regime we build a different surrogate model for $s(μ)$ based on several numerically computed samples.

## 3 Considered Methods

The methods used to build a surrogate model are so-called *supervised* methods. This means that they require a number of input values (input set) and the desired output values for the given inputs. As a baseline method we consider vector function interpolation. In this case a grid for parameter values $μ$ has to be defined. Afterward, we consider machine learning (ML) regression methods. For them the inputs are generated randomly from the uniform distribution. Finally, we take a look at tensor approximation, i.e. cross approximation algorithm. In this case, we also fix a grid for Peclet and Damköhler numbers. Ths is a discrete problem, but the solution can also be extended to the continuous case via interpolation.

### 3.1 Gaussian Process

The first considered method is the Gaussian process regression, one of the machine learning methods. Let us first consider $f(x)$ — a function of a random variable $x$. We suppose that $f$ is the Gaussian process $GP(0, K(x, x'))$ with mean equal to zero, $K(x, x')$ is the kernel function. This means if we take a set $X = \{x_1, \ldots, x_n\}$ of realizations of random variable $x$, then the vector $\boldsymbol{f} = f(X)$ has normal distribution $\mathcal{N}(0, K(X, X))$. Let us suppose that $(X, \boldsymbol{f})$ are the given values (training data) and $X_*$ are the new input values, for which we would like to estimate values of $f$. For the concatenated vector of $\boldsymbol{f} = f(X)$ and $\boldsymbol{f}_* = f(X_*)$ the distribution is the following:

$$\begin{bmatrix} f \\ f_* \end{bmatrix} \sim \mathcal{N}\left(0, \begin{bmatrix} K(X,X) & K(X,X_*) \\ K(X_*,X) & K(X_*,X_*) \end{bmatrix}\right) \tag{6}$$

Then the posterior distribution for $\boldsymbol{f}_* = f(X_*)$ given that $f, X, X_*$ is known:

$$\boldsymbol{f}_*|\boldsymbol{f}, X, X_* \sim N\left(K(X_*,X)K(X,X)^{-1}\boldsymbol{f}, K(X_*,X_*) - K(X_*,X)K(X,X)^{-1}K(X,X_*)\right) \tag{7}$$

The prediction of the model is the posterior mean.

This method can be applied to the prediction of breakthrough curves. The random variable $x$ in this case is a random vector of parameter values $μ$. The function $f$ is replaced by a vector function $\mathbf{s}(μ)$.

### 3.2 Neural networks

Another machine learning method that we consider is the neural network model. In this work we use fully-connected neural networks. This model is represented by a parameterized mapping, composed of linear transforms and nonlinear function $g$:

$$f_θ = L_d \circ g \circ L_{d-1} \circ \ldots \circ g \circ L_1 \tag{8}$$

where $L_i(x; W_i, b_i) = W_i x + b_i$, $W_i \in \mathbb{R}^{h_{i-1} \times h_i}$, $b \in \mathbb{R}^{h_i}$, $θ = \{W_i, h_i\}_{i=1}^d$, $h_0 = 2$, $h_d = N_T$, $h_1 = \cdots = h_{d-1} = h$. $h$ is called the *width* of the network, $d$ is the *depth* of the network, $g$ is the *activation function*. Function $g$ is applied element-wise to each element of the input.

The optimal parameters $θ$ of the network are found by the minimization of the functional called *loss* function. The minimized functional is called the *loss* function. In the experiments we use mean squared error loss. If $X = \{x_1, \ldots, x_n\}$ are the given inputs to the network, $\hat{s}(x)$ is the neural network output for a given $x$, the loss function is written as follows:

$$L(X; \theta) = \sum_{i}^{n} \|\hat{f}(x_i) - f(x_i)\| \tag{9}$$

The loss is usually minimized via a stochastic gradient descent or its variations, e.g. Adam [2]. The optimization procedure is called the *training* of the network.

### 3.1 Cross approximation

The cross approximation method works with tensors. In this case parameter values of interest are located on a fixed grid. The breakthrough curve values for each pair $(Da, Pe)$ on the grid form a 3D tensor $S$. Suppose that breakthrough curve values are known only for a few values $(Da, Pe)$. The task is to reconstruct the whole tensor $S$ from the given samples. To solve this task, we apply the cross approximation algorithm. The algorithm defines the points necessary for the reconstruction and we numerically estimate $s(t)$ values in these points. The method can be further extended from discrete to a continuous case by interpolating over the grid.

Suppose that we have a 3D tensor $S \in \mathbb{R}^{d_1 \times d_2 \times N_T}$. We further assume that $S$ has a low rank structure, i.e. $S$ can be represented in a form:

$$S_{mnt} = \sum_{i=1}^{r_1} \sum_{i=1}^{r_2} G_{ijt} U^1_{im} U^2_{jn}, \tag{10}$$

where $r_1 \ll d_1$, $r_2 \ll d_2$. Let us choose arbitrary indices $k_1$, $k_2$, $A_1 = S[:, k_1, :] \in \mathbb{R}^{d_1 \times N_T}$, $A_2 = S[k_2, :, :] \in \mathbb{R}^{d_2 \times N_T}$. As basis $U^i, i = 1,2$, we will take top $r$ left singular vectors of $A^i$. To find the approximation we need to find the tensor $G$.

To find the tensor $G$ MaxVol algorithm [1] is used. This algorithm looks in the given matrix $A$ for a submatrix of maximum volume ($vol(A) = |detA|$ for a square matrix). This method is applied to matrices $U^1$ and $U^2$. Let us denote the indices of rows in the matrix $U^i$ forming submatrix of the maximum volume as $I^i$, $\hat{U}^i := U^i[I^i]$, $C^i = U^i(\hat{U}^i)^{-1}$. Then $G_{mnt} = \sum_{i=1}^{r_1} \sum_{j=1}^{r_2} (\hat{U}^1)^{-1}_{mi} (\hat{U}^2)^{-1}_{nj} \hat{S}_{ijt}$, where $\hat{S} = S[I_1, I_2, :]$. If all elements in $\hat{S}$ are known, the final approximation is following:

$$S_{ijk} = \sum_{i=1}^{r_1} \sum_{j=1}^{r_2} C^1_{im} C^2_{jn} \hat{S}_{mnk}. \tag{11}$$

## 4 Experiments

The data for training, validation and testing is provided by numerical methods, the usage of breakthrough curves from experiments, or mixed data is straightforward. In total 2000 samples in random positions were sampled. 60% of them were used for training, 20% for validation and 20% for testing. We also generate samples for a $50 \times 50$ uniform grid for tensor approximation. As it was mentioned above, the breakthrough curve is discretized in time. In diffusion dominated case the final time is $T = 1.6$, time step $\Delta t = 6.1 \times 10^{-4}$, $N_T = 2623$, $s \in \mathbb{R}^{2623}$. In the convection dominated case the final time is $T = 0.4$, time step $\Delta t = 6.1 \times 10^{-4}$, $N_T = 657$, $s \in \mathbb{R}^{657}$. To estimate the accuracy of methods we compute two errors. The first one is $l_\infty$ norm of $l_2$ the error between predicted and ground-truth values computed for each sample in the test set:

$$e_1 = \max_{j=1,\ldots,N_{\text{test}}} \sqrt{\frac{\sum_{i=1}^{N_T} (\hat{s}_i(x_j) - s_i(x_j))^2}{\sum_{i=1}^{N_T} s_i^2(x_j)}}, \tag{12}$$

$$e_2 = \sqrt{\frac{\sum_{j=1}^{N_{\text{test}}} \sum_{i=1}^{N_T} (\hat{s}_i(x_j) - s_i(x_j))^2}{\sum_{j=1}^{N_{\text{test}}} \sum_{i=1}^{N_T} s_i^2(x_j)}}, \tag{13}$$

where $s_i(x_j)$ is the prediction of the model for the time step $t_i$ for the test sample $x_j$, $N_{\text{test}}$ is the number of test samples, $x_j$, $j = 1, \ldots, N_{\text{test}}$ is the parameter values for the corresponding sample. We also plot numerically computed and predicted values for each of the methods. The values are provided for the characteristic numbers:
- in diffusion dominated case $Da \in \{0.03, 0.52, 0.98\}$, $Pe \in \{0.03, 0.52, 0.98\}$,
- in convection dominated case: $Da \in \{0.03, 0.52, 0.98\}$, $Pe \in \{6.94, 53.47, 98.06\}$.

These are values from the $50 \times 50$ grid which are inside the considered domain.

## 4.1 1D case

To generate the data required to train machine learning models, we employ an open source library pyMOR [9], to discretize (1)-(5) and to provide suitable input-output mappings, as detailed in [10].

### 4.1.1 Gaussian processes

We take the implementation of Gaussian process regression from the scikit-learn library [4] and compare several kernel functions:

- RBF (Radial Basis Function):

$$K(x_i, x_j) = \exp\left(-d(x_i, x_j)^2/2l^2\right), \tag{14}$$

- Matern:

$$K(x_i, x_j) = \frac{1}{\Gamma(\nu)2^{\nu-1}} \left(\sqrt{2\nu}d(x_i, x_j)/l\right)^l K_\nu\left(\sqrt{2\nu}d(x_i, x_j)/l\right), \tag{15}$$

- Rational quadratic:

$$K(x_i, x_j) = \left(1 + d^2(x_i, x_j)/(2\alpha l^2)\right)^{-\alpha}, \tag{16}$$

where $K_\nu$ is modified Bessel function, $\Gamma(\nu)$ is Gamma function, $d(x_i, x_j)$ is the Euclidean distance function.

As it can be seen from Eq. 7, the memory required to get the solution grows as $N^2$, where $N$ is the number of samples. This problem can be solved by using sparse GP, which we plan to investigate in our future works. Currently, in our experiments 100 training samples were enough to get a sufficiently small error.

The results for diffusion and convection dominated cases are presented in Tables 1 and 2 correspondingly. For particular values of parameters $\mu$ we plot predicted and numerically computed breakthrough curves (Fig. 1). The dotted line denotes the target values, solid line is the prediction. As it can be seen from the plot the predicted values fit well the target values.

| Kernel | $N_{\text{train}}$ | $N_{\text{test}}$ | Fitting time, s | Prediction time, ms | Error $e_1$ | Error $e_2$ |
|---|---|---|---|---|---|---|
| RBF | 100 | 400 | 2.11 | 15.6 | $2.12 \times 10^{-5}$ | $3.85 \times 10^{-6}$ |
| Matern | 100 | 400 | 2.91 | 16.6 | $2.4 \times 10^{-4}$ | $2.81 \times 10^{-5}$ |
| Rational quadratic | 100 | 400 | 0.74 | 13.7 | $1.32 \times 10^{-5}$ | $1.35 \times 10^{-6}$ |

Table 1: Results for Gaussian Processes. Diffusion dominated case.

| Kernel | $N_{\text{train}}$ | $N_{\text{test}}$ | Fitting time, s | Prediction time, ms | Error $e_1$ | Error $e_2$ |
|---|---|---|---|---|---|---|
| RBF | 100 | 400 | 2.18 | 4.92 | 0.0011 | $7.79 \times 10^{-5}$ |
| Matern | 100 | 400 | 0.356 | 13.48 | 0.00305 | $3.6 \times 10^{-4}$ |
| Rational quadratic | 100 | 400 | 0.10.5 | 6.24 | 0.00197 | $1.2 \times 10^{-4}$ |

Table 2: Results for Gaussian processes. Convection dominated case.

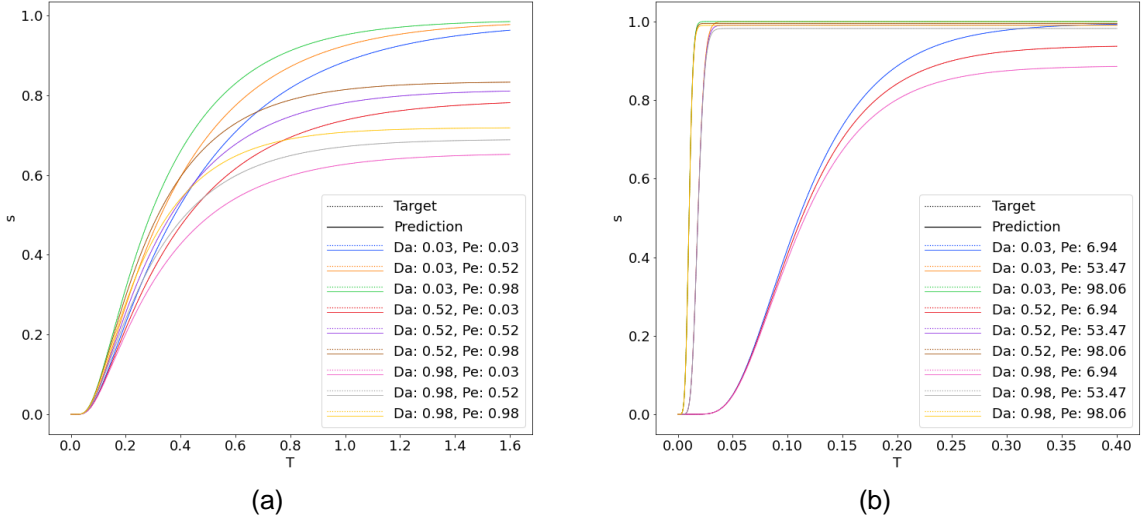

Fig. 1: Breakthrough curves computed numerically and predicted by the Gaussian Process model for a set of Peclet and Damkohler coefficients: (a) – diffusion dominated case, (b) – convection dominated case.

*4.1.2 Neural networks*

The next model, a neural network, we implement using PyTorch [3]. We experiment with different sizes of the network (width $h$ and depth $d$). The used activation function is ReLU ($g(x) = \max(0, x)$). The dataset for the training consists of 1200 samples. For each gradient step optimizer takes one subset of the training dataset or a *batch*. The optimizer iterates over the whole dataset. One such iteration is called an *epoch*. The batch size (number of samples processed at one step) is equal to 128. We also try different sizes of the training set. Validation dataset was used for the stopping criteria: an error on this dataset was estimated and if the error doesn't decrease for 300 epochs then the training was stopped. The learning rate is set at the beginning to $10^{-3}$, when the validation error stops decreasing, the learning rate was multiplied by 0.1 and the training was continued with the same stopping criteria. We update the learning rate only once.

Firstly, we compare the results for different sizes of the network in diffusion dominated case. As it can be seen from the Table 4, increasing the complexity of the model doesn't necessarily improve the performance. So, a wider network of depth 2 has a larger error on the test set. We choose a network of width 100 and depth 3 and plot the predictions of the breakthrough curve for a set of parameters (Fig. 2a). We also use this neural network configuration for the experiments with different sizes of the training dataset. We steadily increase the number of training samples from 100 to 1200. As expected, the test error decreases as the number of samples grows.

We also perform the study on the optimal size of the neural network for convection dominated case (Table 4). Here the deepest network ($d = 10$, $h = 100$) shows the best result. We also provide the comparison plot of predicted and target values of the breakthrough curve for several values of $\mu$

| $h$ | $d$ | # params | $N_{\text{epochs}}$ | | Training time, ms | Prediction time, ms | Error $e_1$ | Error $e_2$ |
|---|---|---|---|---|---|---|---|---|
| | | | $lr = 10^{-3}$ | $lr = 10^{-4}$ | | | | |
| 5000 | 2 | 13.1 M | 400 | 710 | 3637 | 70.7 | 0.0042 | 0.0027 |
| 1000 | 2 | 2.6 M | 510 | 2260 | 1886 | 13.7 | 0.0018 | 0.0008 |
| 100 | 2 | 265.2 M | 2210 | 1070 | 453 | 2.09 | 0.0025 | 0.008 |
| 50 | 2 | 133.9 K | 1360 | 1320 | 353 | 1.91 | 0.006 | 0.0011 |
| 100 | 3 | 275.3 K | 1730 | 2010 | 500 | 2.37 | 0.0039 | 0.0007 |
| 50 | 5 | 141.5 K | 1560 | 1340 | 327 | 2.9 | 0.0056 | 0.0012 |
| 30 | 5 | 84.1 K | 1190 | 570 | 172 | 2.47 | 0.0066 | 0.0015 |
| 10 | 5 | 29.2 K | 1950 | 860 | 232 | 1.82 | 0.0069 | 0.0018 |

Table 3: Results for neural networks of different sizes of the network. Diffusion dominated case.

| $h$ | $d$ | $N_{\text{train}}$ | $N_{\text{test}}$ | $N_{\text{epochs}}$ | | Training time, ms | Error $e_1$ | Error $e_2$ |
|---|---|---|---|---|---|---|---|---|
| | | | | $lr = 10^{-3}$ | $lr = 10^{-4}$ | | | |
| 100 | 2 | 100 | 400 | 3380 | 320 | 52.5 | 0.0066 | 0.0023 |
| 100 | 2 | 300 | 400 | 2470 | 960 | 108 | 0.0066 | 0.0013 |
| 100 | 2 | 600 | 400 | 1980 | 470 | 135.6 | 0.0056 | 0.0012 |

| 100 | 2 | 900  | 400 | 2030 | 750  | 240 | 0.0042 | 0.0009 |
| 100 | 2 | 1200 | 400 | 2210 | 1070 | 453 | 0.0025 | 0.0009 |

Table 4: Results for the neural network for the different number of training samples. Diffusion dominated case.

| $h$ | $d$ | # params | $N_{\text{epochs}}$ | | Training time, ms | Prediction time, ms | Error $e_1$ | Error $e_2$ |
|---|---|---|---|---|---|---|---|---|
| | | | $lr = 10^{-3}$ | $lr = 10^{-4}$ | | | | |
| 100  | 2  | 66.7 K  | 1430 | 6990 | 441 | 0.85 | 0.064  | 0.0115 |
| 1000 | 2  | 660.7 K | 420  | 1020 | 249 | 4.73 | 0.0289 | 0.0182 |
| 100  | 3  | 76.8 K  | 1150 | 600  | 360 | 1.52 | 0.0049 | 0.0017 |
| 100  | 5  | 97 K    | 1670 | 3130 | 477 | 2.1  | 0.007  | 0.0012 |
| 100  | 10 | 174.4 K | 1810 | 1820 | 566 | 2.6  | 0.0071 | 0.0011 |

Table 5: Results for neural networks of different sizes. Convection dominated case.

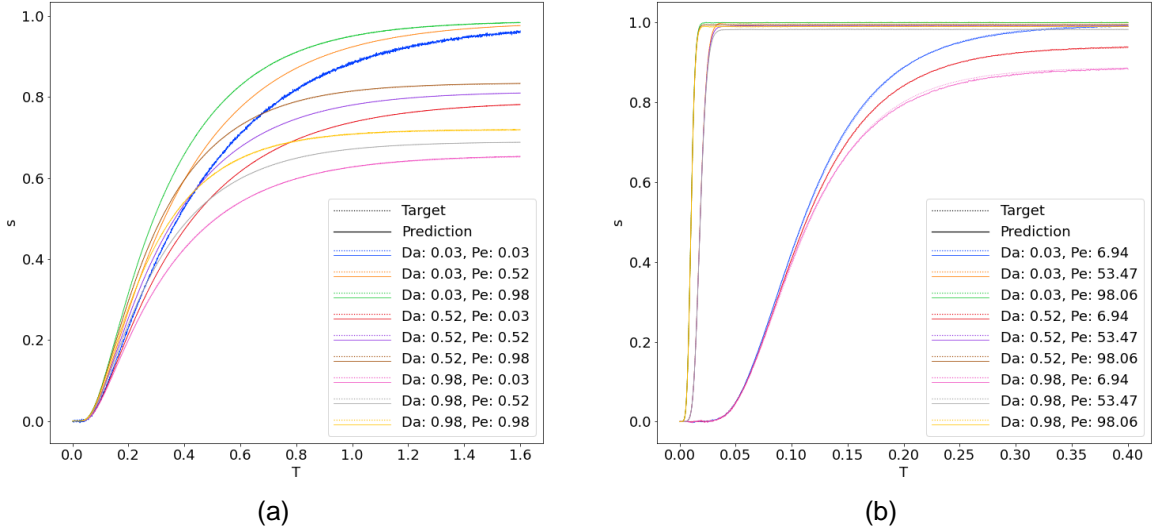

Fig. 2: Breakthrough curves computed numerically and predicted by the neural network model for a set of Peclet and Damkohler coefficients: (a) – diffusion dominated case, (b) – convection dominated case.

### 4.1.3 Cross approximation

Finally, we consider the cross approximation method for tensor approximation. We have a 3D tensor $S \in \mathbb{R}^{50 \times 50 \times N_T}$. We take $k_1 = d_1/2, k_2 = d_2/2$.

In the diffusion dominated case the approximation of rank ($r_1 = r_2 = r$) 4 is enough to achieve error $1.24 \times 10^{-6}$. This means that the breakthrough curve has to be computed for 115 pairs $(Da, Pe)$. The error computed as relative $l_2$ error on positions with missing values (the values are known only for the positions selected by the algorithm). As was mentioned before basis vectors can be interpolated to basis functions, e.g. via polynomials. And then we can compute predictions for arbitrary values of $\mu$. The predictions were computed on the same test set as for Gaussian processes and neural networks and the achieved error ($e_1$) is $1.24 \times 10^{-6}$ for the polynomials of degree 5. The relative $l_2$ error $e_2 = 5.94 \times 10^{-7}$.

In the convection dominated case the approximation of ranks $r_1 = 3$ and $r_2 = 12$ achieves relative $l_2$ error 0.0004 for the samples on positions with missing values. This means that we need 129 precomputed samples to achieve this error. For the polynomial interpolation the $l_\infty$ test error ($e_2$) equals 0.005. The $l_2$ error is: $e_2 = 0.0011$. We used polynomials of degree 5 for the basis functions for $Da$ and 12 for the $Pe$ basis.

The plots of breakthrough curves for several values of $Da$ and $Pe$ on the grid a presented in Figure 3. The times needed to reconstruct tensors, interpolate from basis vectors and to predict values on the test dataset are provided in Table 6. The times averaged over 1000 runs are provided for tensor approximation (reconstruction of tensor $S$), for interpolation to get basis functions and for prediction on the test set.

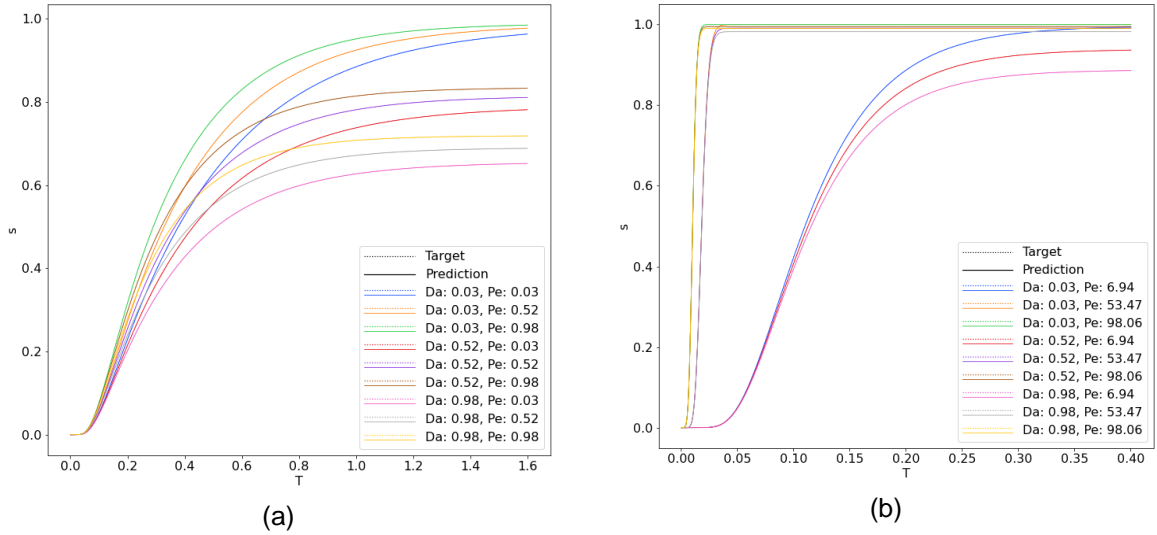

Fig. 3: Breakthrough curves computed numerically and predicted by the cross approximation method for a set of Peclet and Damkohler coefficients: (a) – diffusion dominated case, (b) – convection dominated case.

|  | Regime | Time, ms |
|---|---|---|
| Tensor approximation | Diffusion dominated | $135 \pm 24.6$ |
|  | Convection dominated | $58.8 \pm 3.01$ |
| Interpolation | Diffusion dominated | $0.74 \pm 2.78$ |
|  | Convection dominated | $1.6 \pm 3.23$ |
| Prediction | Diffusion dominated | $18.4 \pm 3.4$ |
|  | Convection dominated | $8.24 \pm 4.35$ |

Table 6: Computing times for cross approximation.

### 4.2 Three-dimensional case

We also perform experiments for a three-dimensional problem for the diffusion-dominated case. We consider a sample of size $50 \times 50 \times 71$. Its geometry is provided in Fig. 4a. Red color denotes inert solid material, green color – the nanoporous active material, yellow and blue inlet and outlet section respectively. The void space is transparent. Slices of the considered velocity field are presented in Fig. 4b. The numerical computations were performed using PoreChem software (see, e.g., [8]).

Here we use the cross approximation algorithm with ranks $r_1 = 2$, $r_2 = 5$. The parameter grid is of size $50 \times 50$, as it was done in the 1D case. For the approximation we take middle slices ($k_1 = k_2 = 25$). This time we do not compute numerically breakthrough curve for the whole grid, so we do not provide the error on the grid. We compute it only for samples in the test set. The examples for the test set are sampled randomly from the uniform distribution, we use interpolation to get the prediction and estimate the error value. The achieved test error ($e_1$) is $0.1111$ for the polynomials of degree $5$, the relative $l_2$ error: $e_2 = 0.0293$. The examples of generated and predicted breakthrough curves are provided in Fig. 5.

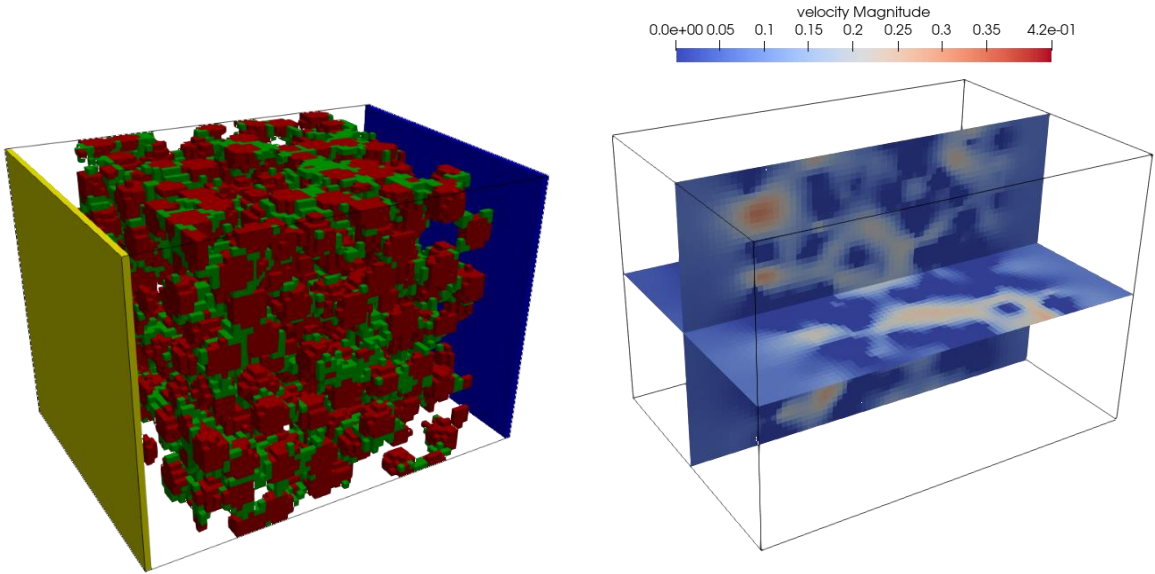

Fig. 4: Geometry in 3D case.

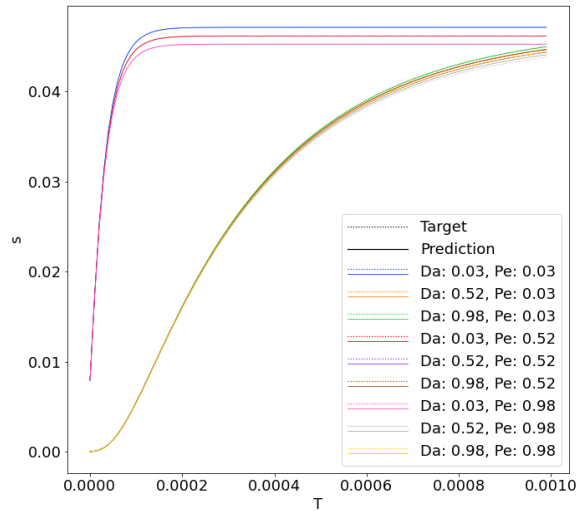

Fig. 5: Breakthrough curves computed numerically and predicted by the cross approximation method for a set of Peclet and Damkohler numbers. 3D case.

## 5  Conclusion

All the considered methods have shown a good performance in predicting breakthrough curves. The cross approximation method has shown very good results. It has achieved an error $e_1 \sim 10^{-6}$ in diffusion dominated case. Such a small error is beyond the needs of the industry, what means that in considering industrial problems much smaller training sets can be used. The drawback of this method is that it requires the low rank assumption for the considered tensor, which has to be checked in particular cases. The neural network in our case is a model with the worst performance with an error $e_1 \sim 10^{-3}$ for the diffusion dominated case. It also needs much more time to train than other approaches. Gaussian process regression has obtained quite a low error ($e_1 \sim 10^{-5}$) and is easy to train. The Gaussian processes however are sensitive to the size of the dataset, as a matrix inversion is needed. This may result in a memory issue. In the convection dominated case all the methods have shown the same order of error values, however, Gaussian processes have got the lowest error and were the fastest to obtain the result. We also present results for cross-approximation in a three-dimensional case, where we could achieve an error $\sim 0.03$. In the future work we plan to perform detailed study of the 3D case for first order reaction, as well as for more complex chemical reactions. Other machine learning methods will be also considered.Embedding the developed machime learning and cross approximation procedures in parameter identification and optimization tasks is also planned.

# References

[1] Goreinov, S. A., Oseledets, I. V., Savostyanov, D. V., Tyrtyshnikov, E. E., and Zamarashkin, N. L.: How to Find a Good Submatrix, Matrix Methods: Theory, Algorithms And Applications: Dedicated to the Memory of Gene Golub. 2010. 247-256.

[2] Kingma, D. P., and Ba, J.. Adam: A method for stochastic optimization, arXiv preprint arXiv:1412.6980, 2014.

[3] Paszke, A., Gross, S., Massa, F., Lerer, A., Bradbury, J., Chanan, G., ... and Chintala, S. (2019). Pytorch: An imperative style, high-performance deep learning library, Advances in neural information processing systems, 2019, 32.

[4] Pedregosa, F., Varoquaux, G., Gramfort, A., Michel, V., Thirion, B., Grisel, O., ... and Duchesnay, E., Scikit-learn: Machine learning in Python, the Journal of Machine Learning Research, 2011, 12, 2825-2830.

[5] Raissi, M., Perdikaris, P., & Karniadakis, G. E., Physics-informed neural networks: A deep learning framework for solving forward and inverse problems involving nonlinear partial differential equations, Journal of Computational Physics, 2019, 378, 686-707.

[6] Sheibani, M., and Ou, G., The development of Gaussian process regression for effective regional post-earthquake building damage inference. Computer-Aided Civil and Infrastructure Engineering, 2021, 36(3), 264-288.

[7] Wang, L., Tao, S., Zhu, P., and Chen, W., Data-driven topology optimization with multiclass microstructures using latent variable Gaussian process. Journal of Mechanical Design, 2021, 143(3)

[8] O. Iliev, P. Toktaliev, On pore scale numerical simulation of complex homogeneous reactions with application to filtration processes, Proceedings of FILTECH 2022 Conference, March 2022.

[9] Milk, R., Rave, S., Schindler, F.: pyMOR – Generic Algorithms and Interfaces for Model Order Reduction, SIAM J. Sci. Comput. 38, 2016, jan, Nr. 5, S194–S216. DOI 10.1137/15m1026614.

[10] Gavrilenko, P., Haasdonk, B., Iliev, O., Ohlberger, M., Schindler, F., Toktaliev, P., Wenzel, T., Youssef, M.: A Full Order, Reduced Order and Machine Learning Model Pipeline for Efficient Prediction of Reactive Flows. Large-Scale Scientific Computing. Springer International Publishing, 2022. – DOI 10.1007/978–3–030–97549–4_43, S. 378–386